**Analyzing Adaptive Scaffolds that Help Students Develop**

**Self-Regulated Learning Behaviors**

Anabil Munshi, Gautam Biswas, Ryan Baker, Jaclyn Ocumpaugh,

Stephen Hutt & Luc Paquette

Abstract

**Background:** Providing adaptive scaffolds to help learners develop self-regulated learning (SRL) processes has been an important goal for intelligent learning environments. Adaptive scaffolding is especially important in open-ended learning environments (OELE), where novice learners often face difficulties in completing their learning tasks.

**Objectives:** This paper presents a systematic framework for adaptive scaffolding in Betty's Brain, a learning-by-teaching OELE for middle school science, where students construct a causal model to teach a virtual agent, generically named Betty. We evaluate the adaptive scaffolding framework and discuss its implications on the development of more effective scaffolds for SRL in OELEs.

**Methods:** We detect key cognitive/metacognitive *inflection points*, i.e., instances where students' behaviors and performance change as they work on their learning tasks. At such inflection points, Mr. Davis (*a mentor agent*) or Betty (*the teachable agent*) provide conversational feedback, focused on strategies to help students become productive learners. We conduct a classroom study with 98 middle schoolers to analyze the impact of adaptive scaffolds on students' learning behaviors and performance.

**Results and Conclusions:** Adaptive scaffolding produced mixed results, with some scaffolds (viz., strategic hints that supported debugging and assessment of causal models) being generally more useful to students than others (viz., encouragement prompts). We also note differences in learning behaviors of High and Low performers after receiving scaffolds. Overall, our findings suggest how adaptive scaffolding in OELEs like Betty's Brain can be further improved to narrow the gap between High and Low performers.

**Implications:** This paper contributes to our understanding of the effectiveness of adaptive scaffolding in OELEs. In addition to improving the scaffolding framework in Betty's Brain, our findings can be used to design/improve adaptive scaffolds to better support SRL behaviors in other OELEs, for instance, by considering affect transitions and cognitive-affective relations to build more effective scaffolds for affect regulation.

## 1. Introduction

An important goal of computer-based learning environments (CBLEs) is to help students develop self-regulated learning (SRL) skills that can make them effective life-long learners (Bransford et al., 2000; Zimmerman and Martinez-Pons, 1990). **Self-regulated learning (SRL)** refers to learners' abilities to understand and control their learning behaviors and environment, to accomplish their learning and problem-solving goals (Panadero, 2017). The SRL process emphasizes the students' autonomy, intrinsic motivation, self-monitoring and control, and self-reflection. Open-ended learning environments (OELEs) have been designed to support SRL development by providing students with (1) *targeted learning goals* (e.g., construct a causal model of a scientific process); (2) *a set of tools* to facilitate the learning and problem-solving processes; and (3) *an open-ended approach* that offers students with choice in how they combine the use of these tools to accomplish their learning goals (Biswas et al., 2016). These OELEs have often used model-building tasks to help students improve strategic thinking skills (Segedy, et al, 2015; Basu et al., 2017; Hutchins et al., 2020).

However, open-ended problem-solving can present significant challenges for novice learners (Kinnebrew et al., 2017; Metcalfe and Finn, 2013), who may have difficulties in using the system tools and explicitly regulate their own learning processes in these environments (Zimmerman, 2002). Still, as researchers have gotten better at tracking and supporting students in their complex learning tasks in OELEs (Azevedo et al., 2010; Biswas et al., 2016; Winne et al., 2010), we have increasing evidence that self-regulation plays an important role in helping students with their problem-solving processes in these environments (Azevedo et al., 2017; Winne, 2017).

Resulting frameworks for studying SRL suggest that researchers should examine an interacting collection of students' *"CAMM"* processes (Azevedo et al., 2017; Bannert et al., 2017), namely:

- *Cognition:* the use of prior knowledge, skills, and strategies to develop solutions for the learning task (Entwistle and Ramsden, 2015);
- *Affect:* the ability to identify and regulate one's emotions during learning (Linnenbrink, 2007);
- *Metacognition:* awareness, monitoring progress toward goals, invoking and applying strategies for effective problem-solving, and periodically reflecting on how to improve performance (Schraw et al., 2006); and
- *Motivation:* the perceived value of the learning task and the subject matter being learned (task value), the self-perceived ability to accomplish the task (self-efficacy) and one's personal goals (intrinsic versus extrinsic) for doing the task (Pintrich, 1999).

Learning environments that scaffold students' CAMM processes can empower them to develop *agency* toward their learning to become more independent and strategic in their learning process (Azevedo et al., 2017; Taub et al., 2020). This form of scaffolding implies *online adaptation*, where the system infers students' behaviors and performance in the OELE and uses this information to provide feedback (Dabbagh and Kitsantas, 2012; Moreno and Mayer, 2000). Plass et al. (2015) discuss adaptive frameworks that provide feedback that is contextualized by the learner's current tasks, intent, and capabilities.

This paper develops and implements a framework for designing and evaluating adaptive scaffolds that support students' self-regulation processes in the Betty's Brain OELE (Biswas et al., 2016; Leelawong and Biswas, 2008). Specifically, our framework includes methods for detecting and understanding students' learning behaviors around key *'inflection points'* when they undergo a change in their cognitive/metacognitive processes. We hypothesize that changes are often linked to their inability to apply productive strategies, thus leading to a drop in their performance. In Betty's Brain, this translates to their inability to correct errors and improve their causal maps. The inflection points provide a basis for generating contextualized *in-the-moment* scaffolds that can help students to become more strategic in their learning and problem-solving tasks. We provide adaptive scaffolding to students through contextualized conversations with two virtual agents: (1) a mentor agent, Mr. Davis, and (2) a teachable agent, Betty.

We evaluate our adaptive scaffolds by conducting a design study and analyzing the effectiveness of our scaffolds using an exploratory data analysis approach. Using the overall differences in students' learning outcomes as a framework, we study how effective our scaffolds are on two groups of students – *High Performers* and *Low Performers*. For each group, we compare students' learning behaviors and performance in the time intervals *before* and *after* they receive a scaffold.

The rest of this paper is organized as follows. *Section 2* reviews prior research on adaptive scaffolding for SRL in computer-based learning environments. *Section 3* discusses Betty's Brain and our previous work on developing adaptive scaffolds. *Section 4* presents our new adaptive scaffolding framework. *Section 5* discusses the empirical methods of a classroom study we conducted to evaluate our scaffolding framework, the findings of which are reported in *Section 6*. Finally, conclusions and implications for future research are provided in *Section 7*.

## 2. Prior Research

Adapting to the specific needs of students has always been a key goal of intelligent computer-based learning environments (CBLEs; Lajoie and Derry, 1993). But novice learners, who are not proficient in using these tools and lack SRL processes, often adopt sub-optimal strategies in their learning and problem solving tasks. This increases the difficulties they face in their learning tasks, and providing students with relevant *in-time* feedback can help them overcome difficulties and become better learners (Puntambekar and Hubscher, 2005, van der Kleij, et al, 2015).

### 2.1 Scaffolding in CBLEs

**Scaffolds** are *"tools, strategies, and guides used to support understanding beyond one's immediate grasp"* (Graesser et al., 2000; Azevedo and Hadwin, 2005). Research has shown that scaffolds and feedback can improve critical thinking (Wood et al., 1976) and learning outcomes, including those for higher-order constructs (Van der Kleij et al., 2015). Scaffolds developed for Betty's Brain — namely contextualized conversational feedback from virtual agents — have led to better overall performance by students (Segedy et al., 2013).

Properly designed scaffolds can help foster self-regulation and engagement and reduce frustration (Lepper and Chabay, 1985; Shute, 2008), but there are sometimes unintended consequences. Students may exploit scaffolding features, as Baker et al. (2004) show in their study of Cognitive Tutors (Koedinger et al., 2006), where students skip learning activity suggestions provided in low-level hints to get to the answers in *bottom-out hints*. In addition, feedback that frequently interrupts workflow or that focuses on summative evaluation can also negatively affect learning (Fedor et al., 2001).

Therefore, a sound design process for adaptive scaffolding should guide students towards the optimal use of scaffold content into their learning and problem-solving processes.

**2.2 Modeling SRL Processes**

Early models of SRL used static *"trait-based"* definitions of the construct (Pintrich et al., 1993; Zimmerman & Martinez-Pons, 1986), but by the late 1990s, this conceptualization shifted toward process-based definitions, including the cyclical phases model (Zimmerman, 2002) and the COPES model (Winne & Hadwin, 1998). The current research consensus on SRL understands it to be a dynamic sequence of cognitive, affective, metacognitive, and motivational (CAMM) events (Azevedo et al., 2017; Panadero et al., 2016).

While SRL models now emphasize *dynamic* processes (Panadero et al., 2016), little research has examined these dynamics in CBLEs, and new methods are needed for detecting and analyzing changes in students' CAMM processes during learning, so that we can develop scaffolds to help students internalize successful SRL processes. In Betty's Brain, prior analysis of interactions between cognitive and affective SRL components showed that virtual agents successfully scaffolded students' learning (Munshi et al., 2018b). In this paper, we extend such earlier findings to design an adaptive scaffolding framework that provides students with agent-initiated (1) guidance on strategies to support their learning tasks and (2) encouragement messages to support their motivation and affect.

**2.3 Designing Adaptive Scaffolds in OELEs**

There have been several approaches to designing adaptive scaffolds in CBLEs. Elsom-Cook (1993) proposed that systems could individualize guidance by varying the form and content of the scaffolds according to the cognitive state of the learner. Later work suggested *"an ongoing diagnosis of the student's current level of understanding of specific and related tasks"* as a pillar of effective scaffold design (Puntambekar and Hubscher, 2005). Basu, et al (2017) demonstrate the effectiveness of providing in-time strategic feedback to students in a computational thinking-based OELE for science learning. In this paper, we build on previous work, and develop adaptive scaffolds that are *strategic* (help students invoke a procedure or piece of knowledge they are unable to apply properly) and informed by students' past learning behaviors and performance. In addition, we provide encouragement scaffolds (praise or reassurance) to help learners avoid or overcome emotions that are detrimental to the learning process.

3**. The Betty's Brain Open Ended Learning Environment**

Betty's Brain, an OELE for middle school science, adopts the *learning-by-teaching* paradigm, where students build *causal models* of scientific processes to "teach" a virtual pedagogical agent, generically named Betty (Biswas et al., 2005; Leelawong and Biswas, 2008). As shown in Figure 1, the system provides students with resources and tools to learn, build, and check their models.

These resources include a **science book**, a set of hypermedia resource pages embedded within the system, that provide the knowledge students need to teach Betty. Students read sections of the book and identify concepts and *causal* (i.e., cause-and-effect) relations between concepts.

An accompanying **teacher's guide** explains procedures students can apply to construct and evaluate their causal maps. Students use the **causal map building tool,** a visual interface with a drag-and-drop menu, to teach Betty. The interface provides students with a visual representation of their causal map encompassing both constructs and subsequent causal links. The menu allows students to add, delete, and modify concepts and links.

Other tools facilitate evaluation of the causal map. The **query and quiz tools** allow students to probe Betty's knowledge of the science concepts and relations, by asking her to take either shorter 'section-specific' (and more targeted) quizzes or a more comprehensive 'Everything' (mastery) quiz. Betty's answers are dynamically generated from the information in the causal map and scored by the **mentor agent**, Mr. Davis. Quiz results direct students to errors in their current causal map, and students can make Betty explain her answers to specific quiz questions by highlighting the links Betty used to answer that question. Effective learners can use this information to make immediate map corrections or to determine which sections of the textbook to read next. Overall, the quizzes help students track Betty's progress, and by implication their own knowledge of the science concepts and relations.

Betty's Brain adopts a socio-constructivist approach to learning that encourages exploration, strategic thinking, and monitoring skills (Biswas et al., 2016). Mr. Davis, the mentor agent, provides relevant strategy-oriented feedback when students have difficulties building and checking their maps. For Mr. Davis to accomplish this, the system must track student progress, but the open-ended nature of the system can make interpreting and adapting to the student quite challenging.

Over the years, researchers have worked to make Betty's Brain more adaptive (Segedy et al., 2013; Kinnebrew et al., 2017). Segedy et al. (2013) used a conversation tree representation (e.g., Adams, 2010) to deliver agent-initiated conversational scaffolds. Biswas et al. (2016) discuss how the listener interface of Betty's Brain facilitates explicit, contextualized communication between the student, Betty, and Mr. Davis by analyzing the current causal map, the most recent quiz results, and the student's recent interactions with the system. However, additional development must consider scaffolding that reflects the changes in the theoretical understanding of SRL (e.g., the shift from SRL as a *static trait* to a *dynamic process*).

## 4. The Adaptive Scaffolding Framework

Our adaptive scaffolding framework builds off the SRL models mentioned in Section 2.2 to support the design and implementation of a set of contextualized conversational feedback constructs in Betty's Brain.

### 4.1 Theoretical Framework

Winne and Hadwin's (1998) COPES model describes self-regulated learners as those who actively manage their learning via enacting and monitoring cognitive and metacognitive strategies. *Cognitive strategies* are typically goal-directed and situation-specific, e.g., read to find a specific piece of information (Weinstein and Meyer, 1994). *Metacognitive strategies* involve more generally applicable processes that include planning, monitoring, and reflecting (Donker et al., 2014; Zhang et al., 2021). While cognitive strategies operate on the knowledge of "objects" or skills (Winne, 1995), *metacognitive* learning strategies involve deliberation on the use of particular cognitive processes and combining them to accomplish larger tasks (Winne and Hadwin, 2008). *Metacognitive monitoring* bridges the gap between cognition and metacognition, as it involves observing and evaluating one's own execution of cognitive processes to control and improve cognition (Kinnebrew et al., 2017).

In this work, we focus on how students monitor and use cognitive and metacognitive strategies. Since novice learners are typically not good at applying, monitoring, and reflecting on their use of strategies, understanding their learning behaviors and possible use of strategies *in context* can help us to design more contextualized scaffolds that better support their strategy development and monitoring process. With this understanding, we also monitor students' affect and performance. By tracking students' progress and interactions with the system, we can identify opportune times (i.e., inflection points) to provide *strategic hints* to help learners become aware of effective strategies for acquisition, construction, and reasoning with knowledge. *Encouragement hints* in the form of praise or reassuring messages can help them regulate their emotions during learning. We believe that contextualized cognitive and metacognitive strategy feedback will help students acquire the necessary SRL processes to become effective and independent learners (Shyr and Chen, 2018).

**4.2 Design**

Designing and delivering strategy-focused feedback in Betty's Brain requires us to consider the different paths that learners may use to accomplish the complex goal of completing a causal map. To be successful, learners must decompose their goal of building a correct map into specific and strategically ordered *tasks* and monitor their progress towards completing their tasks (Winne, 2014). Thus, we need to understand the context of students' actions logged in the Betty's Brain system in terms of the specific goals they are trying to accomplish at that time.

**4.2.1 Detecting Students' Learning Behaviors** *in Context*

Our framework considers three factors to understand the *context* of students' activities and behaviors in Betty's Brain: (1) the current task type (i.e., reading, constructing/refining the causal map, or evaluating the causal map); (2) the student's effectiveness in causal modeling tasks (i.e., adding correct versus incorrect links to their causal map); and (3) the relation between the current task and preceding tasks (i.e., reading a page and then adding links from that page to the causal map, viewing quiz results and then refining certain links related to the results on their map). Taken together, these three factors help us infer students' learning difficulties in different task contexts (e.g., an inability to find the science book pages that contain the information they need to construct causal links, inability to convert the information read into correct causal links, or an inability to analyze quiz results to infer correct versus incorrect links in the map);

To understand, track, and contextualize student behaviors in Betty's Brain, Kinnebrew et al. (2017) developed a hierarchical task model that maps students' tasks and sub-tasks to higher-level (i.e., more general) cognitive processes in the learning environment. A task model decomposes the complex task (i.e., teaching Betty a scientific process by constructing a causal map) into sub-tasks using cognitive task analysis methods (Schraagen et al., 2000). As Figure 2 shows, Betty's Brain requires three cognitive processes: (1) *Information Acquisition* (IA, i.e., reading the hypertext resource pages or taking/organizing notes) (2) *Solution Construction* (SC, i.e., map building/refinement tasks), and (3) *Solution Assessment* (SA, i.e., quiz-related activities). In this paper, we extend our previous model to incorporate an additional task, "Organizing Information" (i.e., taking/editing notes; see Figure 2).

In addition to classifying individual behaviors into these higher-level processes, sequence mining methods can help us derive frequent strategies from logs of students' activities (Kinnebrew et al., 2013) and categorize those that seem strategic (Kinnebrew et al., 2017). For example, when students read resource pages and immediately add to their map, they are demonstrating an *IA $_{(read)}$* → *SC $_{(build\ map)}$* strategy. Such combinations illustrate the coordination and enactment of different cognitive processes, including metacognitive regulation, in the form of learning and problem-solving strategies (Schwartz et al., 2009). For instance, applying an *IA $_{(read\ a\ page)}$* → *SC $_{(add\ correct\ link\ from\ that\ page)}$* strategy shows that a student is able to acquire information from the science book strategically, by (a) identifying the section that contains causal information they need to teach Betty, (b) then interpreting the causal relation from this information correctly, and (c) translating this acquired knowledge of the causal relation into a correct increase/decrease causal link on their map. As students work in Betty's Brain, they may switch between IA, SC, and SA tasks in different ways to accomplish their goals.

The system now uses *pattern detectors* to track students' use of cognitive and metacognitive strategies, which are also analyzed with reference to the resulting changes in causal modeling effectiveness, so these patterns may be classified as productive or unproductive (Munshi et al., 2018b). Prior work has identified a set of productive and unproductive strategies within Betty's Brain (Biswas et al., 2016; Kinnebrew et al., 2017; Munshi et al., 2018a), which we use as the foundation for our adaptive scaffold framework.

**4.2.2 Determining the Conditions for Triggering Scaffolds**

Our new framework contextualizes a (1) *triggering condition* (i.e., a behavior or sequence) to optimize the selection of the (2) the *content of the adaptive scaffold* so that, when a triggering condition is satisfied, the adaptive scaffolding system provides students relevant *in-the-moment* feedback to help them develop effective strategies and become better learners. Specifically, we formalize the selection of *inflection points*—conditions where prior analysis has typically shown a decrease in students' ability to apply effective strategies—as triggering conditions. We argue that these inflection points represent situations when students' self-regulation (CAMM) processes undergo a change as they work on their learning and problem-solving tasks. The infection points suggest *key transitional moments* in students' learning behaviors and productivity. Therefore, the inflection points represent opportune moments for providing *in-the-moment* feedback assistance to struggling students.

Our framework also seeks to address the relationships between cognition and affect (Munshi et al., 2018b) through the inclusion of scaffolds that deliver *encouragement*—either through *reassurance* (e.g., when students find multiple errors in their model after taking a quiz) or *praise* (e.g., when students teach a set of correct causal links to Betty). These help students to manage their affect so that they can continue to engage with the system even when they are struggling.

Table 2 presents a complete list of these scaffolds, along with their inflection point triggering conditions. For example, when a student produces an *IA → SC* sequence with ineffective map-building behaviors (viz., adding incorrect links, deleting correct links, etc.), *in-the-moment* feedback may suggest that the student quiz Betty to assess the effectiveness of their map edits. This may help the student combine *SA→SC* and *IA→SC* (i.e., *SA→IA→SC*) strategies to identify and debug the parts of the causal map by studying the answers to some of the quiz questions. Likewise, an inflection point may reflect key affective experiences. For example, *confusion*, (cognitive disequilibrium) to *frustration* (D'Mello & Graesser, 2012) might occur if Betty's quiz results reflect several ineffective map links. In such situations, triggering affect regulation scaffolding is likely more effective than relying solely on cognitive-metacognitive strategies.

**4.2.3 Providing Conversational Scaffolds at Trigger Conditions**

After identifying inflection points to serve as triggering conditions, we deliver scaffolding through a back-and-forth conversation (Figure 4) between the student and one of the two virtual agents, Mr. Davis or Betty. We have shown that this engages students in more authentic social interactions (Segedy et al., 2013), allowing them to be more active participants in the conversation (D'Mello et al., 2006) since students can direct the discussion toward topics/information they feel are most relevant. The next section discusses our approach for implementing the scaffolding framework in Betty's Brain.

**4.3 Implementation**

An overview of the scaffolding framework implemented in Betty's Brain is shown in Figure 3. Students' primary actions logged in the system, as presented in Table 1, are: (1) *Reading* the resources; (2) *Making notes* as a memory aid and to organize the information read; (3) *Building* and refining the causal map; (4) *Requesting* Betty to take *quizzes*; and (5) *Finding explanations* to quiz answers by checking on the links used to answer questions. These logged actions are mapped onto the higher-level cognitive processes using the task model in Figure 2.

Map-edit activities associated with an increase or decrease in the causal map score (computed as the *number of correct links* − *number of incorrect links* in the map) are identified in the logs by marking them with *-Eff* (effective) and *-Ineff* (ineffective) tags, respectively. For example, an *Edit$_{-Ineff}$* is used for causal map edits that decrease students' map scores. These (*-Eff* and *-Ineff*) labels were also applied to pre-defined task sequences of cognitive and metacognitive strategies, as derived from work by Kinnebrew et al., (2014, 2017), as were labels of *coherence* (i.e., relevant or supported by the information they just received; Segedy et al., 2015).

We applied these labels to data collected from two Betty's Brain classroom studies (March 2017 and Dec 2018), and used two methods to determine candidates for triggering conditions. We used a combination of (1) sequential pattern mining (Kinnebrew et al., 2014) to identify frequent strategies that lead to a decline in performance, and (2) student interviews, to identify times where students articulated difficulties they encountered while working with Betty's Brain. This resulted in nine cognitive/metacognitive inflection points, for which we developed adaptive scaffolds.

Each scaffold (see Table 2) is structured as a *conversation tree* and is delivered to the student by Betty or Mr. Davis. Each step in the conversation is represented by a node in the conversation tree, with opportunities for students to respond at each step. Their responses help guide the subsequent feedback to meet their specific needs. Students have autonomy to exit the feedback at will, controlling the amount of feedback they want based on their judgements of relevance. Two example inflection points are shown in Figure 4 with their corresponding conversation trees: (a) *Edit$_{-Ineff}$* → *Quiz* (when a student edits the causal map incorrectly and then takes a quiz), and (b) *Read-Long* → *Edit$_{-Ineff}$* (when a student spends a long time reading and then edits the causal map incorrectly). In each figure the triggering condition (inflection point) is shown in green, and example conversation text is given in blue.

## 5. Methodology

To evaluate our new scaffolding's effectiveness, we ran a design study in February 2019 with sixth-grade students in an urban public school in the southeastern US. The school's population was 60% White, 25% Black, 9% Asian, and 5% Hispanic, with 8% enrolled in the free/reduced-price lunch program. (Individual classroom demographics were not collected.) During the study, 98 students built a causal model of the human thermoregulation system (regulation of human body temperature, Figure 5) using the updated version of Betty's Brain.

### 5.1 Study Design and Data Collection

The study was conducted over 6 consecutive days. On Day 1, students took a paper-based pre-test that used both multiple-choice & short-answer questions to evaluate students' domain understanding and causal reasoning skills. On Day 2, students worked on a practice unit to familiarize themselves with the Betty's Brain environment. On Days 3-5, students constructed causal models of thermoregulation in Betty's Brain. On Day 6, students took a post-test that was identical to the pre-test.

Betty's Brain logged students' activities and affective states (as detected with trace data) with time stamps as they worked on the system. Specifically, affect detectors captured 5 achievement emotions in 20 second intervals: (1) engaged concentration, (2) boredom, (3) delight, (4) confusion, and (5) frustration using affect detection models (Jiang et al., 2018). All of Mr. Davis' and Betty's conversations were also logged in the system with time stamps. Students' *map scores*, used as a measure of performance, were updated every time students added, deleted, or made changes to their map.

## 5.2 Research Questions for Exploratory Data Analysis

Our primary research objective for analyzing the data collected in this study was to evaluate our adaptive scaffolds, more specifically, to study how the adaptive scaffolds (hints and encouragement prompts) received by students affected their SRL behaviors and performance. To achieve this goal, we used an exploratory data analysis approach that combined students' performance, behaviors, and affect, as logged with time stamps in the learning environment.

First, we studied students' learning outcomes in Betty's Brain. The results (reported in Section 6.1) showed that the study participants could be categorized into two groups (High performers and Low performers) based on their performance differences in Betty's Brain. Therefore, we formulate two primary research questions below, to facilitate a more targeted analysis of the behaviors, and the impact of scaffolds on the behaviors, of each group.

> **RQ 1:** Were there differences in learning behaviors between the High and Low groups as they worked to build their maps in the Betty's Brain environment?

> **RQ 2:** Were there differences in the type and quantity of adaptive scaffolds received by the High and Low group students during the intervention? How did the adaptive scaffolds received by these students affect their learning performance and learning behaviors?

**RQ 1** focuses on the differences in the reading, map editing, and quiz-taking behaviors between the High and Low groups. To answer RQ 1, we study the distribution of time spent by High and Low groups on their primary cognitive activities, and additionally look for differences in the effectiveness and coherence of map building behaviors between the two groups. Then, for a more complete analysis of students' cognitive and metacognitive strategies, we apply differential sequence mining (DSM (Kinnebrew et al., 2013)) to the sequence of activities of the two groups. Section 6.2 discusses the results of these analyses.

**RQ 2** investigates the effectiveness of our scaffolds for the High and Low performing groups. To answer RQ 2, we analyze students' learning strategies and their affective states in Section 6.3 and study the impact that each scaffold (listed in Table 2) had on High and Low groups' map-building performance, learning and map-building behaviors, and their affect.

## 6. Results and Discussion

### 6.1 Learning Outcomes

We operationalize learning outcomes in Betty's Brain using three measures: *(1) Normalized Pre-to-posttest Learning Gains (NLG)*, which provides us with a summative measure of students' learning of the science content; *(2) Final Causal Map Score (FCMS),* which provides us with an overall measure of students' performance in the intervention; and (3) *Current Causal Map Scores (CCMS),* which is a sequence of formative map score measures that update each time a student makes a change to their causal map. The summative score is computed as a learning gain, i.e., $\frac{Post\ score - Pre\ score}{Max\ score - Pre\ score}$, while the map scores are calculated as *# number of correct − # of incorrect causal links* in a student's map after every change they make to their maps.

The distribution of pre-post learning gains was close to normal, with only mild skew and no evidence of kurtosis, justifying use of parametric statistical tests in Table 3. One-way ANOVA tests of the students' summative scores show statistically significant ($p < .05$) learning gains for the science content. Table 3 also shows considerable variation in the summative scores.

Figure 6 shows the distribution of students' final causal map scores, which range from 15 (fully correct map score) max to a min value of -6. (Note that negative numbers represent map scores where the maps have more incorrect links than correct links). We see considerable variation (range $[-6, 15]$; median = 6.6; SD =6.6) (also seen in Kinnebrew et al., 2014).

To better analyze the differences in performance, we divided the students into High ($N = 40$) and Low ($N = 40$) performing groups using a median split on students' final causal map scores. We excluded 18 students who had final causal map scores that were equal to or differed by 1 from the median (i.e., their scores were between $[5, 7]$) from this analysis to create separation between the two groups. As Table 4 shows, while both groups demonstrated significant learning gains, the effect size of *NLG* was much larger for the High group (*Cohen's d* for the High group = 2.28, whereas for the Low group = 0.83) (see Table 5). Therefore, the final causal map score was a consistent predictor of our summative assessment measure.

Next, we next explored the effect of prior knowledge on *FCMS* using 1-way ANOVAs and a 1-way ANCOVA. As Table 5 shows, we used pretest scores as a proxy for prior knowledge, and results showed the group with High final causal map scores also had higher pretest scores ($p < 0.05$; $effect\ size = 0.46$). When we used pretest as a covariate in an ANCOVA on learning gains, we saw an even greater effect size ($p < 0.05$; $effect\ size = 1.56$), suggesting that prior knowledge *alone* does not explain the difference in the learning gains between the two groups. Students' interactions with the system and their strategy use and behaviors in the Betty's Brain intervention also played a role in determining their learning gains. In our continuing analysis of the data, we probed further into students' strategic learning behaviors and the effect of adaptive scaffolds on their behaviors and map building performance.

**6.2 Analysis of Students' Cognitive and Strategic Processes in Betty's Brain**

To answer RQ1, we first compared the differences in time spent on the three primary cognitive processes (IA, SC, and SA) and combined *coherence analysis* (Segedy et al., 2015) with *differential sequence mining* (DSM; Kinnebrew et al., 2013, 2017) to compare the High and Low students' activity patterns and their use of cognitive-metacognitive strategies.

**6.2.1 Cognitive Activity Differences in High and Low Groups**

Using logged data, we computed the proportion of time students spent on the five primary actions related to cognitive processes (IA, SC and SA) in Betty's Brain (see Table 1). Table 6 shows these values for the High and Low groups. Overall, the High group more evenly divided their time between the three cognitive activities ($IA - 27\%, SC - 47\%, and\ SA - 26\%$) than the Low group ($IA - 38\%, SC - 46\%, and\ SA - 16\%$).

The two groups spent a nearly equal proportion of their time editing their causal maps. The High group spent a greater proportion of their time assessing their causal map by taking quizzes and analyzing the results (1.7:1). This difference was significant (*t*-test: $t(78) = 2$, $p < 0.05$). On the other hand, the Low group spent a greater amount of their time reading the science book (1.4: 1; and this difference between the groups was significant, *t*-test: $t(78) = 2.45$, $p < 0.05$), perhaps because of their low prior knowledge. The greater amount of time spent in IA activities may also imply that the Low group had greater difficulty in extracting relevant science knowledge from the resources and translating them into links to build their causal maps.

**6.2.2 Map-building Coherence of High and Low Groups**

We also looked at how the two groups used the information they acquired from reading into causal links on their map (i.e., their *Read→Map-Edit* behavior). We used two measures to contrast the map-building behaviors of the High and Low groups: (1) *coherence* of their Map-Edit actions with prior Read actions; and (2) the *effectiveness* of their Map-Edit actions on their map scores (i.e. whether map scores increased). The High group's *Read→Map-Edit* behaviors were more coherent than the Low group's (88% to 74.8%), but the differences in the coherence measures was not statistically significant ($p > 0.05$). Both groups were engaged in mostly coherent *Read→Map-Edit* behaviors. However, the High group was more effective in adding correct links and correcting incorrect links on their maps than the Low group (63.7% to 45.4%), suggesting that the Hi group was better at identifying relevant content from the science book and translating the content into map building actions that increased their map scores.

**6.2.3 Differences in Use of Strategies between High & Low Groups**

For further analysis, we studied the differences between the frequent activity patterns for the High and Low groups. Kinnebrew, et al (2013) used two measures to find activity patterns between two groups: (1) *s-frequency* (sequence-frequency), which is the proportion of students in each group who used the pattern at least once and (2) *i-frequency* (instance-frequency), which for each group is the average number of times a pattern occurs in a student's sequence. Table 7 shows the *s-frequency* and *i-frequency* values for the set of frequent patterns extracted by the DSM algorithm. In addition, following the notations described in Kinnebrew et al. (2014), we used the tags

*-Eff/-Ineff* to indicate that a student's edit action was effective/ineffective (see Section 4.3). Effective edit actions produced an increase in their map score. The *-Mult* tag was associated with actions that were repeated multiple times in sequence.

Table 7 shows three IA-related activity patterns used frequently by the Low group: (1) *Note → Read-Mult*, (2) *Read → Note,* and (3) *Read-Mult → LinkEdit-Ineff-Mult → Read-Mult*. The first two patterns capture students' reading and note-taking behaviors. Pattern three indicates that the students read multiple pages and added multiple links to their map (but the links added were incorrect), and the students then went back to read activities. This pattern suggests that the Low group was unsuccessful in extracting relevant information from the resource pages and translating them into correct causal links. Further, this shows a lack of awareness on the students' part to use the quiz feature to check their maps, which may have helped them discover incorrect links. As a contrasting example, the *Read-Mult → LinkEdit-Ineff → QuizTaken → LinkEdit-Eff* pattern was used by many more students in the Hi group (ratio of approximately 4:1). This pattern shows how Quiz results may have been used to correct errors in the causal map. Overall, the Low group seemed to lack a *solution assessment* strategy. This is further confirmed by the data reported in Section 6.2.1 that showed that the Low group spent more time reading and editing their maps, and less time in checking their maps using the quiz feature.

In summary, the Low group used ineffective reading and map-building behaviors, and insufficient map-checking behaviors. As discussed, this may be attributed to the Low group's lack of sufficient prior domain knowledge. As a result, they spent more time reading to extract information as compared to the High group. Their reading was ineffective because it led to adding incorrect links to the map.

On the other hand, the High group was more effective in combining their IA and SC activities. Table 7 shows that the High group was about three times more likely to use the (*Read → LinkEdit-Eff-Mult*) strategy. The High group was also four times more likely than the Low group to use the *Read-Mult → LinkEdit-Ineff → QuizTaken → LinkEdit-Eff* pattern. In other words, the High group was also better at using quiz answers to debug their map, and they used the quiz function more often than the Low group (Table 6). They more frequently used the *SA→SC* and *SC→IA* strategies, which they performed twice as often as the Low group.

The analyses in Sections 6.1.1−6.1.3 answers RQ 1 − the High group used more effective and productive strategies for building and debugging their maps, which resulted in higher final map scores and higher learning gains (Table 4).

## 6.3 Impact of Adaptive Scaffolds on Students' SRL Process

To answer RQ 2, we delved deeper into the impact of the scaffolds in Table 2 on students' cognitive processes and their use of strategies. In addition, we also tracked students' affect states and map building performance, especially around the inflection points that triggered the adaptive scaffolds presented to the students.

### 6.3.1 Differences in Scaffolds Received by High and Low Groups

Of the six strategy-related adaptive scaffolds (Hints 1-6), Hint3 and Hint4 were triggered very infrequently for all students ($\leq 5$), so we excluded them from further analyses. For the remaining strategy and encouragement scaffolds, we computed: (1) the average number of times students in the High and Low groups received a scaffold; and (2) the number of times students in each group received each scaffold during the intervention. Table 8 lists the number of times (0 to 4+) an adaptive scaffold was received.

Table 8 shows that the High group received more feedback than the Low group. For three of the seven adaptive scaffolds, Hints 2, 5, and 6, this difference was statistically significant, with t(43)=1.6 for Hint2, t(76)=2.3 for Hint5, t(78)=4.7 for Hint6, and $p < .05$ in each case. This result seems counter-intuitive because one would expect that low performing students should receive more adaptive scaffolding to help them overcome their difficulties. However, as we discussed above, the triggering conditions for a number of these scaffolds (see Table 2) required students to take quizzes to assess their progress. Since the High performers took quizzes more often than the Low performers (Table 6), they received hints more often than the Low performers. This distinct difference in the number of hints received by the two groups may imply that the High group benefited from receiving more feedback than the Low group. In the next section, we investigate whether they benefited more from the help they received.

### 6.3.2 Impact of Scaffolds on the High & Low Groups

To answer the second part of RQ 2, i.e., how the adaptive scaffolds affected students' learning performance and behaviors, we tracked the change in their performance, their related cognitive and strategic processes, and their affect after they received feedback from Mr. Davis or Betty. For this temporal analysis, we created sequences of *scaffold-triggered 'before' and 'after' intervals*, where the *after* interval for an adaptive scaffold started just after the adaptive scaffold was given to the student and extended till the student received the next scaffold from the system. Similarly, the *before* interval started from when students received the last adaptive scaffold to when the current scaffold was provided. To illustrate this, we consider a student who received two adaptive scaffolds during the course of their learning session – a Hint2 at time $t_i$ and a Hint5 at time $t_j$. For the Hint2 scaffold, the student's before interval was $[0, t_i]$ and after interval was $[t_i, t_j]$, where the time 0 represents the start of the current session. Similarly, for Hint5, the before interval was $[t_i, t_j]$ and after interval was $[t_j, end]$, where $end$ represents the end time of the session.

For each scaffold, we studied students' causal modeling performance, behaviors and emotions in the *before* and *after* intervals to determine the effectiveness of the scaffold. We used *"average map-score slope"* as a measure of their causal modeling performance in a *before* or *after* interval (Kinnebrew et al., 2014). Map-score slope in an interval is calculated as the slope of a regression line fitted to a student's map scores as a function of their map edits in that interval over time. By analyzing the Map-score slope and the students' strategic behaviors in the intervals before and after they received each scaffold, we analyzed the effectiveness of the scaffolds on students' performance and learning behaviors over time.

Next, we discuss our findings of the impact of the different Hint and Enc scaffolds on High and Low students' learning behaviors and performance.

**Hint1 (Mark Correct Links on map):** This hint reminded students to *mark the correct causal links on their map* so they could keep track of the correct links on their map. This hint was triggered when the student took a quiz in which at least one of the answers was graded correct. Mr. Davis delivered this feedback if a student did not follow up by marking the links associated with correct answers as correct on their map. Table 8 shows that 55% of the High group ($n = 22$) and 80% of the Low group ($n = 32$) did not receive this hint. The remaining 18 High and 8 Low students received the hint once or twice during the entire intervention. Since many students did not mark their links, the trigger condition for this hint may need to be revised to ensure that more students receive the hint and use it productively.

*Behavior*: For the 18 High and 8 Low students that got this hint at least once, we study if students adopted this link marking activity, and if it helped them to improve their learning behaviors and performance. In the interval before receiving Hint1, only one High student and one Low student had marked at least one link each on their maps. In the interval after they received Hint1 for the first time, 23 links were marked by students (8 by the High students and 15 by the Low students). Within the High group, 13 of the 18 students did not mark any links after getting the hint, 4 students marked 1 link each, and 1 student marked 4 links on their map. The student who marked the 4 successive links followed these actions by deleting an incorrect link from their map, suggesting that keeping track of correct links may have aided their debugging process.

For the Low group, 4 of the 8 students who got the hint did not mark any links, one student marked 1 link, one student marked 2 links, one marked 4 links, and one marked 8 links on their map upon receiving the hint. The student who marked 8 links switched between looking at the quiz results and marking the correct links and then deleted two incorrect links from the map, suggesting that this student was systematically applying this hint and marking the correct links also helped the student identify incorrect links that needed to be deleted from the map. Four High students and 2 Low students got Hint1 a second time during their learning session, but none of these students marked any links following the second time they received the hint.

*Discussion*: In the current study, marking correct links may have had a marginally positive effect on students' ability to keep track of their correct links. We will have to improve the feedback and provide additional information to help students understand the advantages of marking links correctly. In past studies, we have seen students correct links on their map, but later delete/change the link when some of the other quiz answers are incorrect (Kinnebrew et al., 2013). Therefore, marking links may be a useful memory aid to ensure correct links are not deleted or changed to be incorrect.

**Hint2 (Assess map by taking Quiz):** This adaptive scaffold was designed to inform students that having Betty take a quiz is an effective strategy to assess the correctness and completeness of their maps. Betty delivered this hint to encourage students to check on how much she was learning. The hint was triggered when students read multiple pages in the science book but added incorrect links to the map. Table 8 shows that 28 High and 17 Low students received the hint at least once. A few High students received the hint up to seven times and one Low student received the hint four times. Twelve students from the High group and 23 students from the Low group never received this hint.

*Behavior*: We study the impact of the hint on students' relevant cognitive behaviors, i.e., taking quizzes and then assessing the quiz results by viewing the answers and checking the explanations. We also look at map-building performance changes from *before* to *after* they got this adaptive scaffold. 28 High and 17 Low students got Hint2 at least once, but only 12 High students and two Low students had taken a quiz before they received Hint2. After receiving Hint2 for the *first time*, 24 of the 28 High students and 12 of the 17 Low students took a quiz. Three of the High students and one Low student took the quiz multiple times. When students got Hint2 a second time, they took a quiz immediately upon receiving the hint.

This suggests that the majority of the students who received Hint2 responded to the agent's feedback by taking a quiz, but it is not clear that they internalized this assessment strategy and used it on their own as they progressed in their map building activities. We get more insights into the impact of Hint2 on students' map assessment behaviors by studying their activities after they got Hint2 and took a quiz. Of the 24 High students who took a quiz after getting Hint2 for the first time, 9 students went on to view the causal explanations to specific quiz answers, suggesting that these students were engaged in extended map assessment behaviors by analyzing the correct and incorrect answers in their quiz. For the Low group, only two students viewed quiz explanations after they received the hint for the first time, but the numbers increased upon getting the hint a second time. Unlike the 9 High group students, the Low group students did not engage in deeper map assessment behaviors the first time they received the hint. Over time, more Low students started analyzing quiz behaviors more extensively.

*Performance*: The average map-score slope in the interval before Hint2 was $-0.02$ for the High group and $-0.29$ for the Low group, suggesting the students were not performing well in their map-building activities. After students received Hint2 for the first time, the average map-score slope for the High group increased to $0.45$, but the average map-score slope for the Low group decreased further to $-0.45$. This implies that the High group was more effective in using the feedback to assess and correct errors in their maps than the Low group, who had difficulties in assessing and correcting errors in their maps. However, when the Low students received Hint2 multiple times, their after-hint map-score slope increased, and students who received Hint2 a third time achieved an average slope of $0.33$ in the after phase. This finding suggests that it took multiple hints for the Low group to develop an effective map assessment strategy.

*Discussion*: Overall, Hint2 was effective for both groups. However, the High performers were more adept at using the explanations for analyzing quiz answers to improve their map-building performance. In contrast, it took multiple hints for the Low group to develop an effective strategy using Hint2. This suggests that more details on how and why Hint2 is useful may help the Low students develop effective debugging strategies faster.

**Hint5 (Debug from Map) and Hint6 (Debug from Read).** Hint5 and Hint6 were both designed to have Mr. Davis make more detailed suggestions on how students could debug the errors in their causal map after they had taken a quiz. Hint5 pointed them to specific erroneous links on the map (SC), whereas Hint6 focused on going back and reading specific pages in the science book to find information to correct their erroneous links (IA). Table 8 shows that students received Hint5 (trigger: *SC-Ineff→SA*) and Hint6 (trigger: *SA→IA (multiple reads)*) more often than the other scaffolds. High students received Hint5 14 times on average and Hint6 24 times on average. Low students received Hint5 9 times on average and Hint6 14 times on average. There could be two reasons why these two hints dominate: (1) the use of less stringent filtering criteria imposed by the pattern detectors as triggers for these hints (see Section 4.2.1); and (2) students frequent inability to use the quiz results effectively. Section 6.2.3 showed that the High group used this this strategy more often than the Lo group. In other words, they were using quizzes in an attempt to debug their maps more often than the Low group, and therefore, they got the quiz-triggered hints more often. One may argue that the Low performers needed this hint more to help them develop effective debugging strategies to correct errors in their map. In the future, we may need to take into account students' performance and their current cognitive abilities in terms of their map checking behaviors in specifying the hint triggering conditions to better match student needs. We may also have to suggest to Low performers who check their maps very infrequently to take quizzes more frequently. If they do and do not demonstrate good map debugging strategies, they would receive Hint5 and Hint6 to help them use the quiz results to debug their maps.

Hints 5 and 6 were often delivered in succession (38% of the time they received either hint) because both hints originated from quiz-taking episodes Therefore, we studied the impact of these two hints for three different cases: (a) when students received *Hint5 only*, (b) when students received *Hint6 only*, and (c) when students received both *Hint5 and Hint6 in succession*.

**Hint5 only.** Hint5 was designed to help students focus on the incorrect quiz answers and to figure out which links needed to be fixed to correct the answer. Table 8 shows that all High group students received Hint5 at least once, with 37 students (92.5%) getting the hint four times or more during their learning session. 38 of the 40 students in the Low group, got Hint5 at least once, with 29 students (72.5%) receiving this scaffold four times or more during the intervention.

*Behavior*: In the interval *before* receiving Hint5, the High group spent 58% of their time and the Low group 50% of their time on map-building activities. *After* receiving Hint5 for the first time, the High group spent an average of 57% of their time on map edits. This number increased to 64% after the third time they received the hint. For the Low group, the map editing time increased from 58% after the first time to 81% (a significant increase) after the third time they got the hint. Therefore, as their causal maps became more complex, Hint5 seemed to have a greater impact on students' map-building efforts, especially for the Low group.

*Performance*: After receiving Hint5 for the first time, the average map-slope score changed from 0.16 to 0.14 for the High group and from $-0.23$ to 0.18 for the Low group. This suggests a marked improvement in performance for the Low group. Receiving this hint more than once had a positive effect on both groups, with the net value of the average map-slope score after getting Hint5 being 0.2 for the High group and 0.36 for the Low group. This suggests that the students used the information provided in Hint5 to successfully find and correct incorrect links on their map.

**Hint6 only:**. All students in the High and Low groups received Hint6 at least once during the intervention. 39 High students and 36 Low students got this hint four times or more.

***Behavior***: Before receiving Hint6, the High group spent 37% of their time and the Low group spent 44% of their time reading the science book. After receiving Hint6 for the first time, the High group spent 38% of their time reading, while the Low group, who were already reading more than the High group, spent 54% of their time on the reading task. The time allocated to reading by the High group did not change much after receiving Hint6 multiple times. For the Low group, the reading time was the highest (54%) after the first time they received Hint6, and decreased thereafter to a stable value in the range 31−37% after receiving the hint three or more times. This suggests that the High group, who were better at finding information in the science book, did not have to devote additional time to reading after they received Hint6, but they probably were more strategic in their approach (Table 7). By contrast, the Low group spent more time reading after they got the hint for the first time. The change in map-slope score from *before* to *after* the hint gives us more insight on whether the Low students were able to use this additional strategic reading to debug their maps.

***Performance***: The average map-slope for the High group was 0.14 before they received Hint6 and did not change significantly upon receiving the hint, suggesting that Hint6 by itself did not result in a performance change for this group. For the Low group, the map-slope score changed from −0.07 in the interval before the hint to an average of −0.2 after the hint, with the score dropping to −0.43 as they got additional hints even though they read more. Despite students increasing their reading time after receiving the hint (especially the first time they received it), the Low group did not become more effective readers. As discussed earlier, this may be attributed to their low prior knowledge or their inability to extract relevant knowledge when reading the science book text. This will be investigated further in future work.

**Hints5+6:** When students received the two hints in quick succession, they always received Hint5 before they received Hint6, prompting us to label these situations as Hints5+6. There was no significant change in student activities or performance trends after receiving Hints5+6. Their *before* and *after* interval map slopes showed fluctuations across intervals instead of a uniform upward/downward trend. On further inspection, we found that different students even within the same group, and at different points during their learning session, reacted differently to receiving Hints5+6, with some resorting to more reading and others performing more map editing activities, but overall, there were no substantial differences in behavior between the High and Low groups. We believe that students became confused upon receiving the two hints in succession. hence their non-uniform reactions to the scaffolding. Emotion likelihood values generated by the affect detectors (Jiang et al., 2018) provide circumstantial evidence for this belief. Confusion scores increased after receiving Hints5+6; for the High group it went from 8% to 13% and for the Low group it went from 8% to 12%. This suggests a need for a "minimum inter-scaffold time" or more explicit information why Hint6 is being given following Hint5 to give students opportunities to better understand how to use the two hints.

**Encouragement Prompts**: The three encouragement prompts provided (Table 2) were: (1) *Praise*, delivered by Betty to commend students when her quiz score improved because they had added correct links to their maps; (2) *Praise + Quiz*, where Mr. Davis praises the student for teaching Betty well by adding links to the map, and then suggesting that the student ask Betty to take a quiz to check if her performance is improving; and (3) *Encouragement (Reassure)*, delivered by Betty when students make errors (by adding incorrect links or deleting correct links) in their map and take a quiz, which shows that Betty is not getting better at answering quiz questions. Unlike the other two encouragement scaffolds whose purpose was to praise students for making progress, Enc3 (Reassure) was intended to keep students from getting demotivated when they saw quiz answers graded as 'incorrect' on their map.

None of the encouragement hints had any impact on students' affective states or their performance; therefore, we do not discuss them in detail. While students did not show any negative transitions in their affective states after the feedback, they also did not show any positive changes. In a complex open-ended learning environment like Betty's Brain, it is possible that the reassurance would have been more useful if it was associated with actionable (strategic) information that the student could use to improve their current maps (Tan and Biswas, 2006). We need to redesign our encouragement scaffolds in view of the above findings to make them more useful towards improving students' affective experiences in the learning environment.

## 7. Conclusions and Future Work

In this paper, we have designed and implemented an adaptive scaffolding framework to help students develop and refine their cognitive and metacognitive behaviors in the Betty's Brain learning environment. Our system continually monitors students' map building performance and the activities they perform in the environment. A deeper understanding of their activity sequences linked to their performance, led to the definition of inflection points and corresponding triggers to generate in-time feedback for students in the form of conversational hints. Results from a study run in a $6^{th}$-grade classroom showed that the students achieved significant pre- to post-test learning gains. However, we also observed large differences in learning outcomes, so we grouped students into high performers (*High*) and low performers (*Low*) using a median split on their final causal map scores to better understand how students' performance and learning behaviors may be related, and how effective the hints were for each of these groups.

The High group showed a higher level of prior domain knowledge than the Low group, which may have resulted in the High students spending significantly less time in knowledge acquisition (reading) and more time on solution assessment (taking and interpreting quizzes). Overall, the High group was better at applying learning strategies. For instance, while both High and Low groups were coherent in their *Information Acquisition (IA)→Solution Construction (SC)* process, the High group was more adept at extracting the correct causal links from the science book, while the Low group had trouble identifying correct relations from reading (possibly in part due to their lack of prior knowledge). The Low group also lacked a good map assessment strategy. On the other hand, the High students successfully used the *Solution Assessment (SA)→ Solution Construction (SC)* strategy, i.e., they used quiz results to identify and correct errors in their maps.

The High group's success in building more correct causal models may be attributed to them engaging in a combination of self-regulated learning strategies. These strategies include the ability to identify and extract relevant information from the science book and add them as correct links to their maps *(IA→SC)*. A second strategy is the ability to monitor their progress by taking timely quizzes, and assessing quiz results *(SA→SC)* to correct errors in their causal models. Lower prior knowledge and the inability to apply SRL strategies during model building and debugging seem to be important reasons in differentiating the Low performers from the High performers in Betty's Brain. Adaptive in-time scaffolding produced mixed results: it helped the High performers more than it helped the Low performers. We expected that the Low performers needed more frequent feedback initially to learn effective strategies, but our trigger conditions need to be improved to support this process more effectively.

Our adaptive scaffolding framework implemented in Betty's Brain was designed to identify moments when students had learning difficulties and respond with strategic hints and encouragement messages to help them adopt more effective SRL strategies. The findings reported in Section 6.3.2 show that some of our scaffolds were useful for students whereas some others did not serve their intended purpose or need to be refined to make them more effective. For example, feedback suggesting link annotation by marking correct links was found to have a marginal positive effect on students' ability to correct links, and may be improved in the future by further emphasizing to learners how link annotation can help them spot incorrect links in their map more easily. The difference in High and Low students' behaviors after they take a quiz based on the mentor's suggestion also presents opportunities for improving the feedback for Low performers, by providing additional scaffolding to help them interpret their quiz results and develop more effective $SA{\rightarrow}SC$ strategies. In fact, scaffolds that helped Low performers with map debugging after they took a quiz were successful in improving their causal modeling performance.

In addition, our results show that the three types of encouragement feedback provided by the system were largely ineffective. Detecting students' affect transitions during learning can help us offer more meaningful scaffolds tailored to support the regulation of negative emotions, such as frustration and boredom (D'Mello and Graesser, 2012). The use of affect detectors, such as developed by Jiang et al. (2018) to track students affective states, and combining that affective information with students' recent activities and performance in the system can help to generate feedback in the context that addresses students' cognitive and metacognitive processes along with suggestions for emotion regulation, when appropriate. Additionally, affective feedback that provides encouragement, especially praise for doing well, is more useful when the learner considers

MY APA DOCUMENT                                                                                           38

the praise to be credible (O'Leary et al., 1977). Therefore, it will be helpful to combine encouragement hints with suggestions for learners to develop strategic behaviors. We may point out to students how similar strategies they used in the past helped them debug errors and improve their maps.

More generally, this study and past studies (e.g., Leelawong & Biswas (2008); Schwartz, et al (2009); Roscoe, et al (2013)) demonstrates that in-time adaptive scaffolding directed toward learning and applying SRL strategies while problem solving in OELEs is essential to help students become better learners. However, the adaptive scaffold triggers need to be designed to be more cognizant of students' needs, and the scaffolding content may need to take into account the prior knowledge of Low performers. An important role of adaptive scaffolding is to close the gap between High and Low performers, and careful design considerations and additional empirical analyses are needed to discover how to narrow this gap.

Overall, this paper has developed an effective design and implementation of an adaptive scaffolding framework in Betty's Brain. In the future, we hope to refine and generalize the in-time adaptive scaffolding framework to better support students' SRL behaviors in the Betty's Brain and in other OELE environments.

MY APA DOCUMENT                                                                                                        41Devolder, A., van Braak, J. and Tondeur, J. (2012) Supporting self-regulated learning in computer-based learning en-vironments: systematic review of effects of scaffolding in the domain of science education. Journal of Computer Assisted Learning, 28, 557–573.

DMello, S. and Graesser, A. (2012) Dynamics of affective states during complex learning. Learning and Instruc-tion, 22(2), 145–157.

D'Mello, S. K., Craig, S. D., Sullins, J. and Graesser, A. C. (2006) Predicting affective states expressed through an emote-aloud procedure from autotutor's mixed-initiative dialogue. International Journal of Artificial Intelligence in Education, 16, 3–28.

Donker, A., de Boer, H., Kostons, D., van Ewijk, C. and van der Werf, M. (2014) Effectiveness of learning strategy in-struction on academic performance: A meta-analysis. Educational Research Review, 11, 1–26.

D'Mello, S., Lehman, B., Pekrun, R. and Graesser, A. (2014) Confusion can be beneficial for learning. Learning and Instruction, 29, 153–170.

Elsom-Cook, M. (1993) Student modelling in intelligent tutoring systems. Artif. Intell. Rev., 7(3–4), 227–240. Entwistle, N. and Ramsden, P. (2015) Understanding student learning (routledge revivals). Routledge.

Fedor, D., Davis, W., Maslyn, J. and Mathieson, K. (2001) Performance improvement efforts in response to negative feedback: the roles of source power and recipient self-esteem. J. Manag., 27(1), 79–97.

Graesser, A., Wiemer-Hastings, K., Wiemer-Hastings, P., Kruez, R. and Group, T. T. R. (2000) Autotutor: A simulation of a human tutor. Journal of Cognitive Systems Research, 1, 35–51.

MY APA DOCUMENT                                                                                           46Schraagen, J. M., Chipman, S. F. and Shalin, V. L. (2000) Cognitive task analysis. Psychology Press.

Schraw, G., Crippen, K. J. and Hartley, K. (2006) Promoting self-regulation in science education: Metacognition as part of a broader perspective on learning. Research in science education, 36, 111–139.

Schwartz, D. L., Chase, C., Chin, D. B., Oppezzo, M., Kwong, H., Okita, S., Biswas, G., Roscoe, R., Jeong, H. and Wag-ster, J. (2009) Interactive metacognition: Monitoring and regulating a teachable agent. Handbook of metacog-nition in education, 340–358.

Segedy, J., Kinnebrew, J. and Biswas, G. (2013) The effect of contextualized conversational feedback in a complex open-ended learning environment. Educational Technology Research and Development, 61(1), 71–89.

Segedy, J. R., Kinnebrew, J. S. and Biswas, G. (2015) Using coherence analysis to characterize self-regulated learning behaviours in open-ended learning environments. Journal of Learning Analytics, 2, 13–48.

Self, J. (1988) Student models: what use are they. Artificial intelligence tools in education, 73–86. Shute, V. (2008) Focus on formative feedback. Rev. Educ. Res., 78(1), 153–189.

Shyr, W.-J. and Chen, C.-H. (2018) Designing a technology-enhanced flipped learning system to facilitate stu-dents' self- regulation and performance. Journal of Computer assisted learning, 34, 53–62.

Skipper, Y. and Douglas, K. (2012) Is no praise good praise? effects of positive feedback on children's and university students' responses to subsequent failures. British Journal of Educational Psychology, 82, 327–339.

# Tables

**Table 1:** Student Activities and Cognitive Processes Associated with Learning in Betty's Brain

| Activity | Description | Cognitive Process |
|---|---|---|
| Read | Student reads resource pages (to learn about domain) or teacher's guide (to get suggestions for teaching Betty) | IA |
| Make Notes | Student takes/edits notes generated from reading resources for organizing information and for future reference | IA |
| Causal Map Edits | Student adds/deletes concepts or adds/deletes/modifies links to build/refine their causal map | SC |
| Take Quiz | Student asks Betty to take a quiz on a topic and reviews quiz results | SA |
| Quiz Expl | Student probes deeper into quiz results by checking the causal links Betty used to answer specific quiz questions | SA |

**Table 2:** Inflection point triggers and their corresponding scaffolds

*(a) When the trigger condition is related to unproductive/ineffective activities*

| Inflection Point Trigger | | Corresponding Scaffold | |
|---|---|---|---|
| *Task/Activity Context* | | *Scaffold Type* | *Content Overview & Excerpts* |
| Information acquisition (Read-Long) → Ineffective Solution construction (Edit-Ineff) | | Strategic hint: Assess by Quiz **Hint2** | Betty suggests taking a quiz, as a good assessment strategy to help debug errors in the map. *"Hi, I think you just added a causal link on your map after looking at the science book. ... Do you think I am ready for a quiz now?"* |
| Ineffective Solution construction (Edit-Ineff) → Solution assessment (Quiz) | Case 1: AND The student has not marked the recently edited incorrect links. | Strategic hint: Mark Wrong **Hint3** | Mr. Davis suggests marking the possibly incorrect links on map as "could be wrong", as an efficient map organization strategy. *" From the quiz results, looks like Betty may have some incorrect links on her map. You can mark those links as 'could be wrong'. Do you want to know more? ..."* |
| | Case 2:<br>WHERE The Edit-Ineff was a *shortcut link* addition (e.g., an A→C link instead of an A→B→C link) | Strategic hint:<br>Shortcut Link<br>**Hint4** | Mr. Davis explains how to identify & correct shortcut links.<br>*"From the quiz, it seems you may have an incorrect shortcut link on your map. Do you want to know more about shortcut links? ..."* |
| | Case 3<br>(No additional contexts) | Strategic hint:<br>Debug from<br>Map **Hint5** | Mr. Davis provides map debugging strategies<br>to fix model errors identified from quizzes, progressing from high-level feedback to more specific corrective hints.<br>*"One of the links going out of 'hypothalamus response' is wrong. Try to find out which one it is."* |
| | Case 4<br>(No additional contexts) | Encouragement:<br>Reassure<br>**Enc3** | Betty provides an encouragement message<br>to ensure that the student is not demotivated after seeing their errors in the quiz results.<br>*"... Sometimes I find all this a little tricky. But with you to teach me, I'm sure we can do it."* |
| Solution assessment (Quiz) → Information acquisition (Read-Long) | | Strategic hint:<br>Debug from<br>Read **Hint6** | Mr. Davis provides progressive hints to support reading the pages relevant to map errors,<br>as an efficient map debugging strategy.<br>*"You are missing a link that comes out of 'heat loss.'. Try reading up on Page 'Response 1: Skin Contraction' and see if you can find the link."* |

*(b) When the trigger condition is related to productive/effective activities*

| Inflection Point Trigger | Provided Scaffold | |
|---|---|---|
| *Task/Activity Context* | *Scaffold Type* | *Content Overview & Excerpts* |
| Information acquisition (Read-Long) → Efficient Solution construction (Edit-Eff) | Encouragement:<br>Praise & Quiz<br>**Enc2** | Mr. Davis praises the student for teaching her well, and suggests taking a quiz to find evidence for their teaching progress.<br>*"Looks like you're doing a good job teaching correct causal links to Betty. Make sure you check her progress by asking her to take a quiz"* |
| | Strategic hint: | Mr. Davis suggests marking the possibly correct links on the map as "correct", as an efficient map organization strategy. |

| | | | |
|---|---|---|---|
| Efficient Solution | Case 1 | Mark Correct | *"If Betty got an answer graded correct, remember to mark those links as 'correct' in the map. This can help you keep track of what you have taught her correctly so far. Do you know how to …"* |
| construction (Edit-Eff) → Solution assessment (Quiz) | | **Hint1** | Betty praises the student for doing a good job of teaching her an efficient causal model. *"Wow! I think I have some correct links on the map. This is fun! Thanks, A."* taught her correctly so far. Do you know how to …" |
| | Case 2 | Encouragement: Praise **Enc1** | Betty praises the student for doing a good job of teaching her an efficient causal model. *"Wow! I think I have some correct links on the map. This is fun! Thanks, A."* |

**Table 3:** Pre-post learning outcomes: All students (n=98)

| Pre/post question type | Pre-test score | Post-test score | Pre to post learning gains (normalized) | Pre to post 1-way ANOVA | Effect size |
|---|---|---|---|---|---|
| | *mean (sd)* | *mean (sd)* | *mean (sd)* | *F-ratio (p-value)* | *Cohen's d* |
| **Multiple Choice (Max=8)** | 2.73 (1.3) | 4.7 (1.92) | 0.35 (0.41) | 66 (< 0.05) | 1.2 |
| **Short Answer (Max=15)** | 0.86 (1.03) | 2.82 (2.33) | 0.14 (0.15) | 56 (< 0.05) | 1.09 |
| **Overall (Max=23)** | 3.59 (1.9) | 7.52 (3.9) | 0.2 (0.19) | 80 (< 0.05) | 1.28 |

**Table 4:** Learning outcomes: High and Low groups

| | Pre-post outcomes | | | | | Final Model score |
|---|---|---|---|---|---|---|
| Category | Pre score Max=23 | Post score Max=23 | Learning Gain (NLG) | Pre vs post 1-way ANOVA | Effect size ANOVA | Final map score (FCMS) |
| | *mean (sd)* | *mean (sd)* | *mean (sd)* | *F-ratio (p-value)* | *Cohen's d* | *mean (sd)* |
| **High (n=40)** | 4.19 (2.06) | 10.45 (3.3) | 0.33 (0.17) | 100.92 (0.0001) | 2.28 | 13.55 (−0.12) |
| **Low (n=40)** | 3.32 (1.65) | 5.14 (2.77) | 0.1 (0.12) | 13.28 (0.0005) | 0.83 | −0.12 (2.78) |

**Table 5:** Significance of Prior Knowledge & Learning Gain Differences between High & Low groups

| Hi & Low NLG Differences | | Hi & Low Prior Knowledge Differences | | Hi & Low NLG Differences, Accounting for Prior Knowledge Differences | |
|---|---|---|---|---|---|
| 1-way ANOVA on Learning Gain | | 1-way ANOVA on Pre-test Scores | | 1-way ANCOVA on Learning Gain | |
| Hi vs Lo | | Hi vs Lo | | Hi vs Lo | |
| F-ratio (p-value) | Effect Size *Cohen's d* | F-ratio (p-value) | Effect Size *Cohen's d* | F-ratio (p-value) | Effect Size *Cohen's d* |
| 45.82 ($< 0.05$) | 1.53 | 4.15 ($< 0.05$) | 0.46 | 47.47 ($< 0.05$) | 1.56 |

**Table 6:** Distribution of primary cognitive activities of High and Low groups

| Category | Percentage of total time spent in each cognitive activity | | | | |
|---|---|---|---|---|---|
| | Information acquisition (IA) | | Solution construction (SC) | Solution assessment (SA) | |
| | Read | Make Notes | Causal Map Edits | Take Quiz | Quiz Expl |
| **High (n=40)** | 26.2 | 0.5 | 47 | 23.13 | 3.2 |
| **Low (n=40)** | 37.3 | 0.7 | 46.1 | 14.1 | 1.8 |

**Table 7:** DSM Results: High versus Low (Max gap = 1; s-threshold=0.5)

| Pattern | I-Frequency Statistics | | | | S-Frequency Statistics | | |
|---|---|---|---|---|---|---|---|
| | i-Support Hi | I-Support Lo | t-test (p-value) | Effect Size (Cohen's f) | s-Support Hi | s-Support Lo | s-Frequent Group |
| *Note → Read-Mult* | 0.5 | 1.65 | 0.003 | 0.7 | 0.28 | 0.6 | Lo |
| *Read → Note* | 0.68 | 1.6 | 0.028 | 0.5 | 0.35 | 0.53 | Lo |
| *Read-Mult → LinkEdit-Ineff-Mult → Read-Mult* | 0.38 | 1.15 | 0.001 | 0.75 | 0.3 | 0.6 | Lo |
| *Read → LinkEdit-Eff-Mult* | 1.58 | 0.58 | 0.000 | 0.91 | 0.8 | 0.43 | Hi |
| *Read-Mult → LinkEdit-Ineff → Quiz-Taken → LinkEdit-Eff* | 1.63 | 0.4 | 0.001 | 0.78 | 0.83 | 0.2 | Hi |
| *Quiz-Mult → Read-Mult* | 9.93 | 5.55 | 0.000 | 0.91 | 1 | 0.98 | Both |

**Table 8:** Number of adaptive scaffolds for students in the High (n=40) and Low (n=40) groups

| Adaptive Scaffold | Category | No. of times a student got the scaffold | | No. of students (% of category) who got the scaffold | | | | |
|---|---|---|---|---|---|---|---|---|
| | | Range | Mean (SD) | never | 1 time | 2 times | 3 times | 4+ times |
| Hint1 Mark Correct | Hi | 0-2 | 1.2 (0.4) | 22 (55%) | 14 (35%) | 4 (10%) | 0 | 0 |
| | Lo | 0-2 | 1.3 (0.4) | 32 (80%) | 6 (15%) | 2 (5%) | 0 | 0 |
| Hint2 Assess by Quiz | Hi | 0-7 | 2.3 (1.5) | 12 (30%) | 10 (25%) | 9 (22.5%) | 5 (12.5%) | 4 (10%) |
| | Lo | 0-4 | 1.6 (0.9) | 23 (57.5%) | 10 (25%) | 4 (10%) | 2 (5%) | 1 (2.5%) |
| Hint5 Debug from Map | Hi | 1-35 | 13.8 (8.2) | 0 | 2 (5%) | 0 | 1 (2.5%) | 37 (92.5%) |
| | Lo | 0-37 | 9.5 (7.4) | 2 (5%) | 3 (7.5%) | 4 (10%) | 2 (5%) | 29 (72.5%) |
| Hint6 Debug from Read | Hi | 3-45 | 23.8 (13.8) | 0 | 0 | 0 | 1 (2.5%) | 39 (97.5%) |
| | Lo | 1-43 | 9.7 (8.8) | 0 | 1 (2.5%) | 2 (5%) | 1 (2.5%) | 36 (90%) |
| Enc1 Praise | Hi | 0-4 | 1.5 (0.8) | 18 (45%) | 14 (35%) | 6 (15%) | 1 (2.5%) | 1 (2.5%) |
| | Lo | 0-2 | 1.1 (0.3) | 32 (80%) | 7 (17.5%) | 1 (2.5%) | 0 | 0 |
| Enc2 Praise & Quiz | Hi | 0-4 | 1.6 (1) | 22 (55%) | 12 (30%) | 2 (5%) | 3 (7.5%) | 1 (2.5%) |
| | Lo | 0-2 | 1.2 (0.4) | 30 (75%) | 8 (20%) | 2 (5%) | 0 | 0 |
| Enc3 Reassure | Hi | 0-3 | 1.3 (0.6) | 28 (70%) | 9 (22.5%) | 2 (5%) | 1 (2.5%) | 0 |
| | Lo | 0-3 | 1.4 (0.7) | 32 (80%) | 6 (15%) | 1 (2.5%) | 1 (2.5%) | 0 |

# Figures

**Figure 1:** System interfaces for the Betty's Brain learning environment

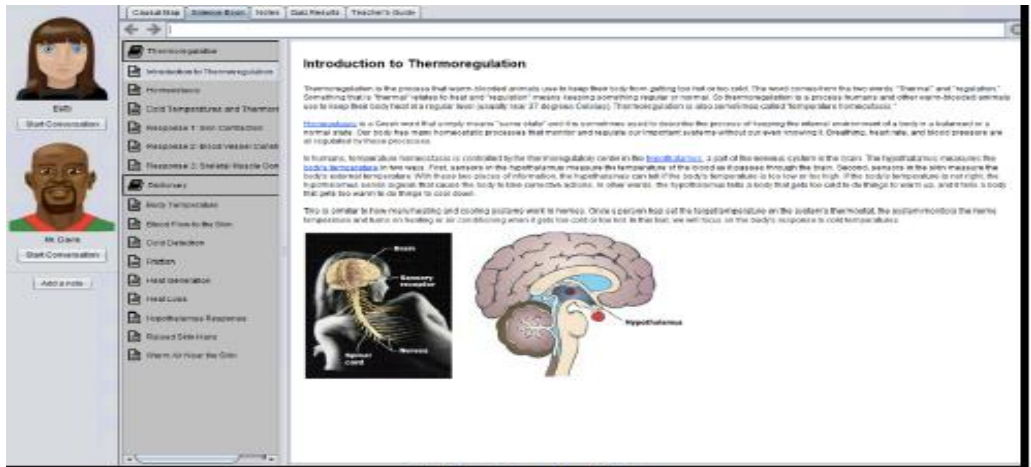
(a) The 'science book' view

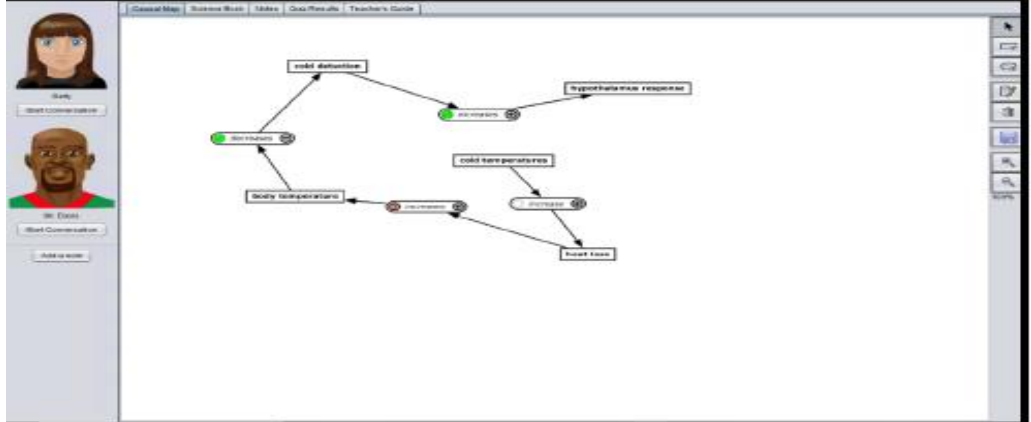
(b) The causal map view

(c) The quiz results view

(a) The 'science book' view, (b) The 'causal map' view, (c) The 'quiz results' view

**Figure 2:** The Hierarchical Task Model for the Betty's Brain Environment (Modified from Kinnebrew et al., 2017)

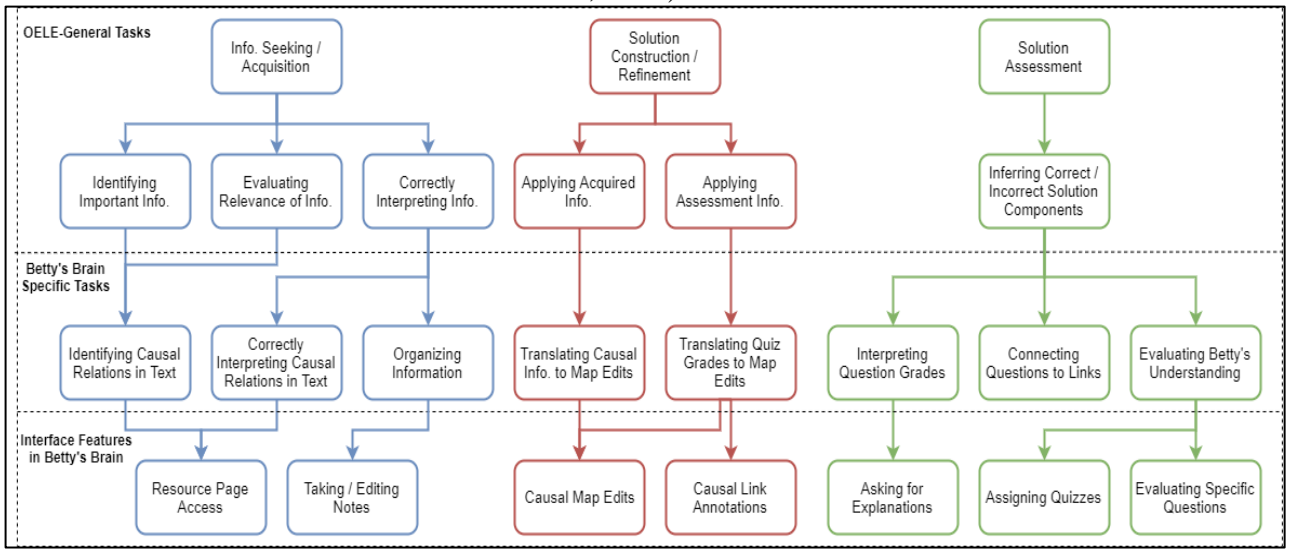

**Figure 3:** Implementation of the adaptive scaffolding framework in Betty's Brain

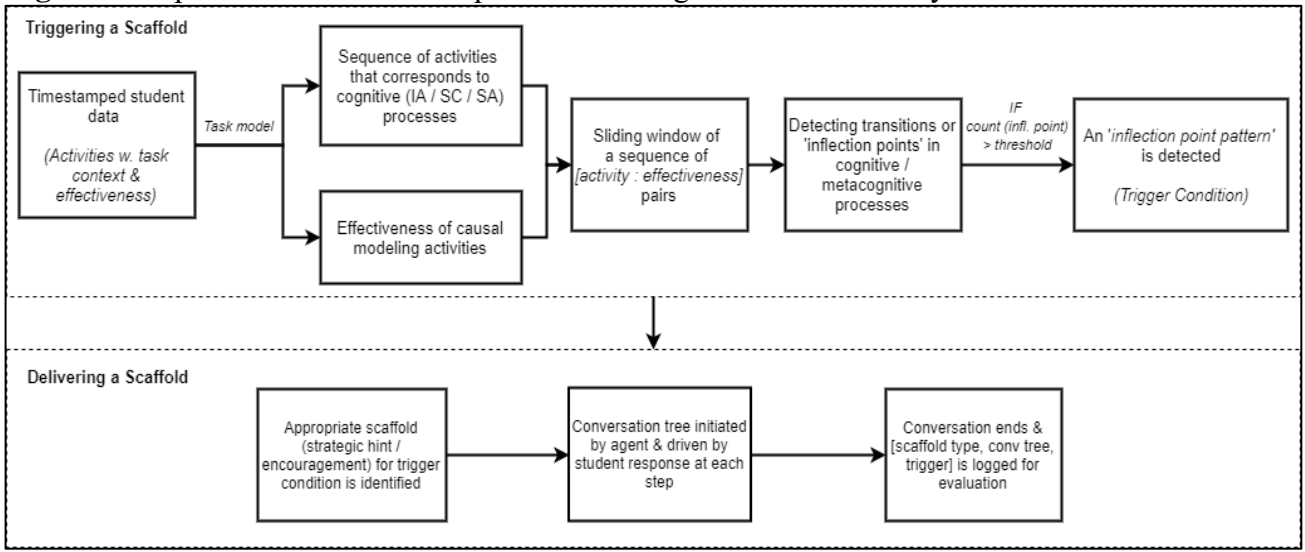

**Figure 4:** Conversation tree representations of two scaffolds from our framework
(a) Progression levels of a conversation tree for a map-debugging scaffold by Mr. Davis

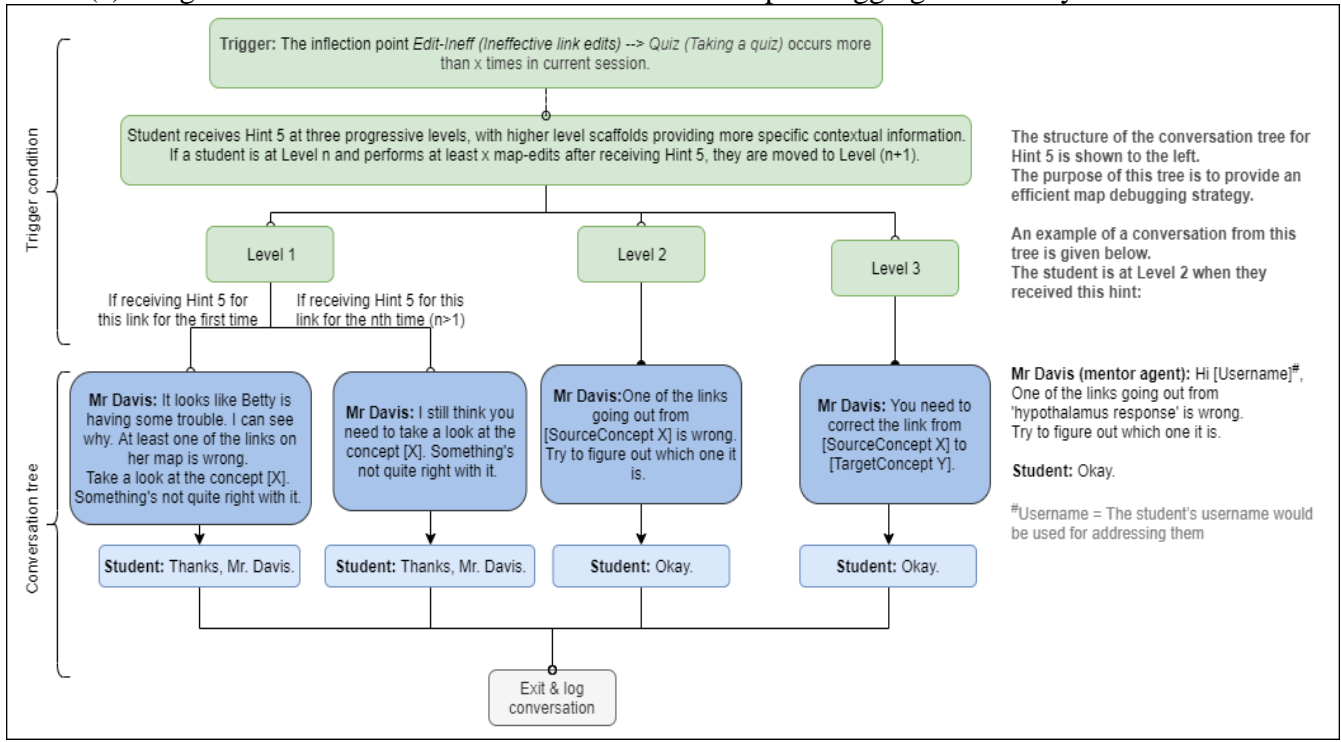

(a) Conversation tree for Hint2, a map-assessment scaffold initiated by Betty

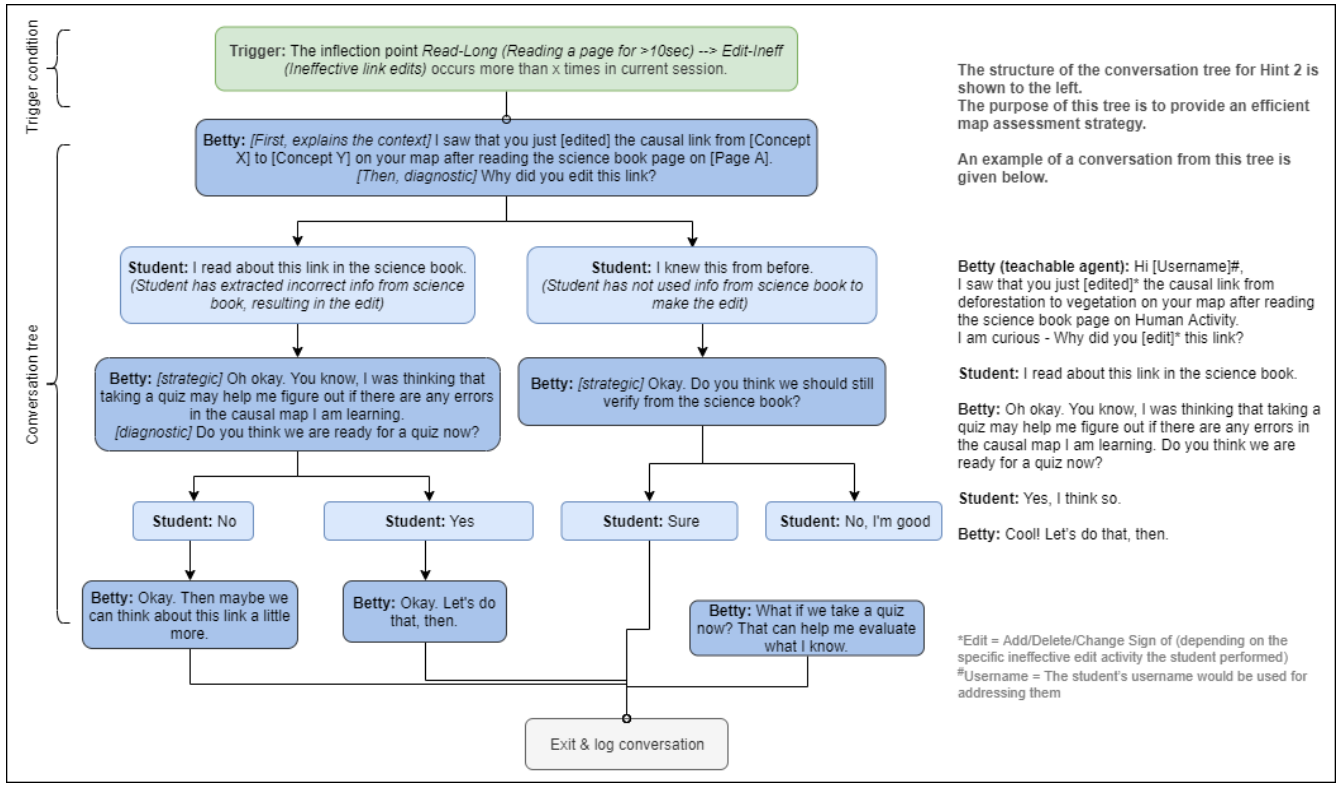

**Figure 5:** Causal map of the human thermoregulation process in Betty's Brain

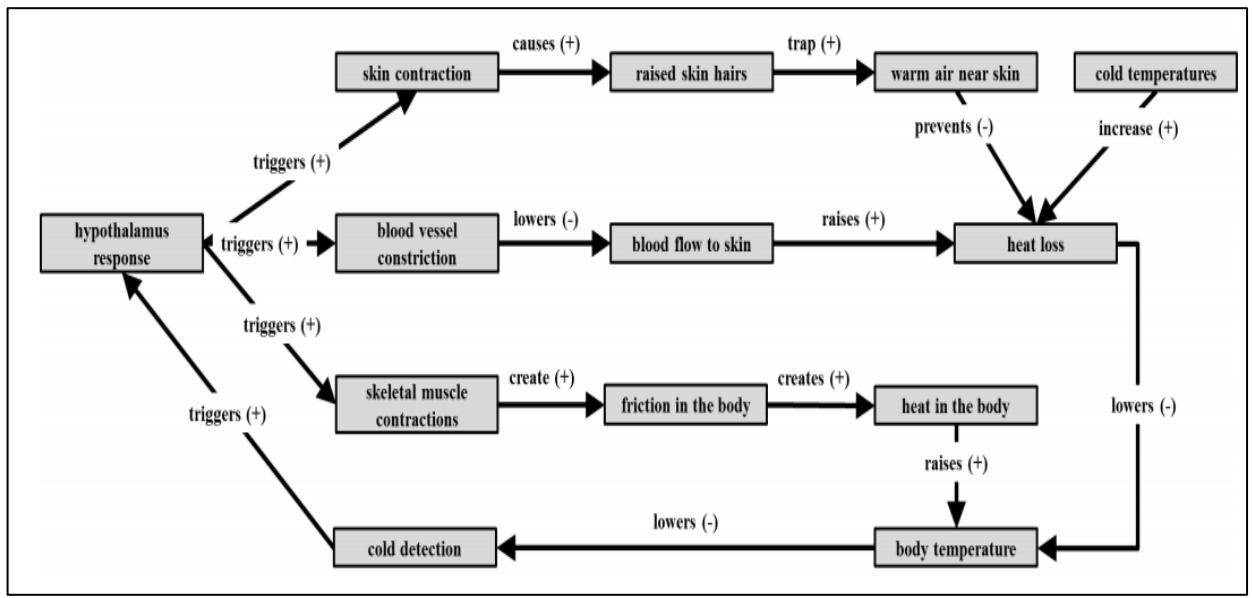

**Figure 6:** Distribution of the final causal map scores (x-axis), by number of participants who achieved the score (y-axis), in the empirical study

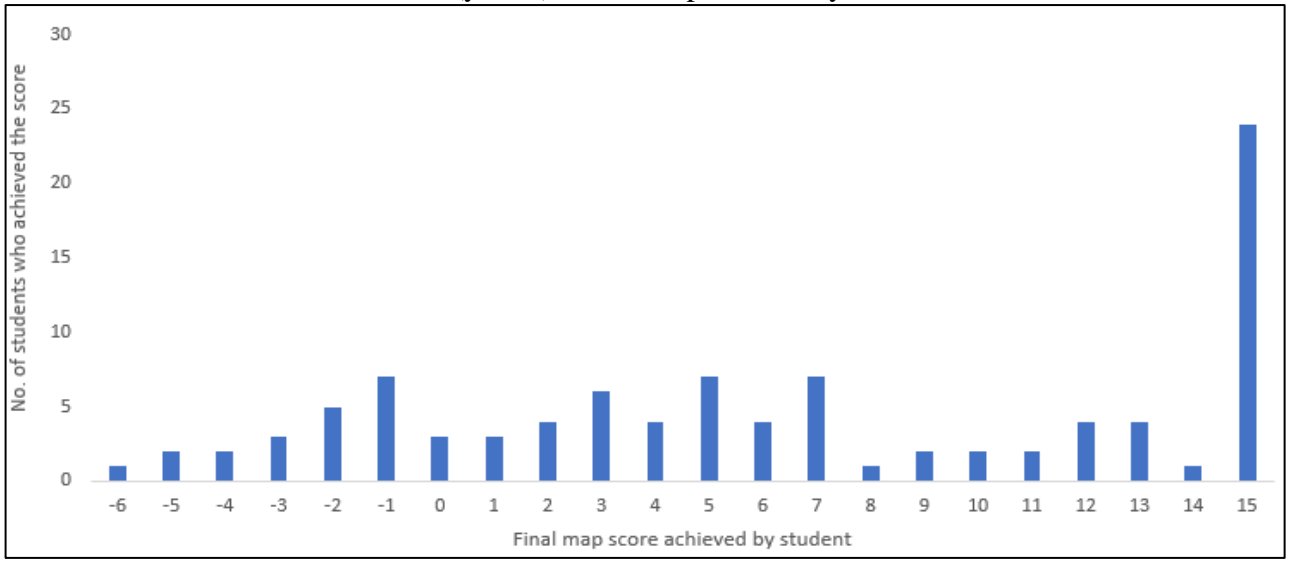